\documentclass[aps,pre,10pt,superscriptaddress,nofootinbib,twocolumn,amsmath,amssymb,final,letterpaper]{revtex4-1}
\usepackage{graphicx}
\usepackage{subfigure}
\usepackage{times}
\usepackage{bm}
\usepackage{xcolor}

\usepackage[colorlinks=true,linkcolor=blue,urlcolor=blue,citecolor=blue,anchorcolor=blue]{hyperref}

\hypersetup{pdfdisplaydoctitle=true,pdftitle={Exact analytical solution of binary dynamics on networks}}
\hypersetup{pdfauthor={E. Laurence, J.-G. Young, S. Melnik and L.J. Dubé}}
\hypersetup{pdfsubject={Exact analytical solution of binary dynamics on networks}}

\newcommand{\redcopy}[1]{#1}

\begin{document}
\title{Exact analytical solution of irreversible binary dynamics on networks}

\author{Edward Laurence}
\email{edward.laurence.1@ulaval.ca}
\affiliation{D\'epartement de Physique, de G\'enie Physique, et d'Optique, Universit\'e Laval, Qu\'ebec (Qu{\'e}bec), Canada G1V 0A6}
\author{Jean-Gabriel Young}
\affiliation{D\'epartement de Physique, de G\'enie Physique, et d'Optique, Universit\'e Laval, Qu\'ebec (Qu{\'e}bec), Canada G1V 0A6}
\author{Sergey Melnik}
\affiliation{MACSI, Department of Mathematics \& Statistics, University of Limerick, Ireland}
\author{Louis J. Dub\'e}
\email{ljd@phy.ulaval.ca}
\affiliation{D\'epartement de Physique, de G\'enie Physique, et d'Optique, Universit\'e Laval, Qu\'ebec (Qu{\'e}bec), Canada G1V 0A6}

%
\begin{abstract}
In binary cascade dynamics, the nodes of a graph are in one of two possible states (inactive, active), and nodes in the inactive state make an irreversible transition to the active state,
as soon as their precursors satisfy a predetermined condition.
We introduce a set of recursive equations to compute the probability of reaching any final state,  given an initial state, and a specification of the transition probability function of each node.
Because the naive recursive approach for solving these equations takes factorial time in the number of nodes, we also introduce an accelerated algorithm, built around a breath-first search procedure.
This algorithm solves the equations as efficiently as possible, in exponential time.
\end{abstract}
\maketitle

\section{Introduction}
\label{sec:introduction}

Irreversible binary-state dynamics model rapid information transmission in complex systems.
In these dynamics, nodes are in one of two states (inactive, active), and nodes in the inactive state make an irreversible transition to the active state, as soon as their precursors satisfy a predetermined condition.
In a popular formulation of the problem \cite{Gleeson2008}, transition conditions are expressed by a set of response functions $\{F_i(m)\}_{i=1,..,N}$ that give the probabilities that nodes will make a transition from an inactive to an active state, based on the number $m$ of their active precursors.
This general formulation, known as \emph{cascade dynamics}, encompasses multiple important dynamics as special cases \cite{Gleeson2008}.
Noteworthy examples are site and bond percolation \cite{Broadbent1957,Callaway2000}, the Watts model of threshold dynamics \cite{Watts2002}, and \redcopy{Susceptible-Infected (SI) models} ~\cite{Newman2002}.
As a result, cascade dynamics have relevant applications in fields as diverse as epidemiology \cite{Meyers2007,Newman2002}, economics \cite{Haldane2011}, and neuroscience \cite{Zhou2015,Kaiser2007,Kaiser2010}.

Cascade dynamics present a difficult mathematical challenge: Predicting its outcome on arbitrary network topologies is notoriously hard.
Only in some special cases do we know of analytical solutions that are both simple and elegant.
Perhaps the most famous special case is that of ensembles of tree-like networks \cite{Newman2001,Molloy1995,Molloy1998}, for which probability generating functions (PGF) \cite{Wilf1990,Callaway2000,Gleeson2008,Gleeson2013,Melnik2014}, and message passing (MP) \cite{Shrestha2014,Karrer2014} methods yield exact analytical predictions of the important observables of the dynamics  (e.g., size of the giant component, critical propagation threshold, etc.).
In fact, their predictions are so accurate that these methods are routinely applied to real networks, even though the underlying hypotheses are no longer valid; in many cases, this leads to suprisingly good approximations of the true outcome  \cite{Melnik2011,lokhov2015dynamic}. 
However in many important cases, experiment and theory are at odds.
It is now understood that these discrepancies can---in part---be traced back to existence of local correlations in real systems, whose effects are  overlooked by tree-based theory  \cite{Melnik2011}.

There are a few promising methods---developed within the general framework of cascade dynamics---that address this issue by including local correlations using, e.g., a fixed prevalence of cliques on a tree-like backbone \cite{Hackett2011,Hackett2013} or with fixed degree-degree correlations \cite{Melnik2014}.
However, the most versatile proposition to date comes from percolation theory \cite{Allard2012b,Allard2015,Karrer2010}; it consists in mixing PGF methods with exact solutions on arbitrary motifs.
The general idea behind this method is simple: One first decomposes a network in small local subgraphs (motifs), then solves percolation on these graphs, and finally combines the local solutions to obtain a global solution \cite{Allard2015}.

There appears to be no conceptual obstacles to adapting this method to general cascade dynamics---the clique-based methods of Refs.~\cite{Hackett2011,Hackett2013} are in fact special cases of the general approach.
There is, however, a major practical bottleneck: Exactly solving cascade dynamics on small graphs is, at best, tedious \cite{Hackett2013}.
This bottleneck becomes a barrier when motifs are too diverse, or too large.
In Refs.~\cite{Allard2012a,Allard2015}, a costly but systematic algorithm is introduced to handle enumeration and averaging of the traversal probabilities in the \emph{percolation problem}.
The goal of the present paper is to delineate the equivalent enumeration algorithm in the much more general context of \emph{cascade dynamics}.

The paper is organized as follows.
We first define cascade dynamics, and obtain recursive equations for the probability of every outcome, on arbitrary network topologies and general cascades (Sec.~\ref{sec:formalism}).
We then discuss the practical aspect of the formalism in Sec.~\ref{sec:numerical_methods}, i.e., how to compute the solution of the recursive equations as efficiently as possible.\footnote{We also provide a reference implementation of our solver at \url{github.com/laurencee9/exact_binary_dynamics}.}
In Sec.~\ref{sec:results}, we illustrate the power of the formalism in a case-study of complicated mixed dynamics occurring on small directed networks.
We gather our conclusions and perspectives in Sec.~\ref{sec:conclusion}.
Three appendices follow. In the first, we prove that our method is equivalent to  the analogous method derived in the context of percolation theory (Appendix~\ref{Sec:AppendixBondPercolation}). In the second, we work out an explicit example (Appendix~\ref{Sec:AppendixExample}). In the third and last Appendix, we give a detailed calculations of the worst-case computational complexity of our algorithms (Appendix~\ref{Sec:AppendixComplexity}).

\section{Cascade dynamics on arbitrary graphs}
\label{sec:formalism}
\subsection{Definition of the dynamics}
\label{subsec:definition}
We consider an irreversible binary-state dynamics occurring on an arbitrary network of $N$ nodes.
This dynamic process is well defined on graphs that contain directed edges \cite{Gleeson2008}, and we therefore encode the structure of the network in a binary-valued, potentially asymmetric $N \times N$ adjacency matrix $\bm{A}$.
We adopt the convention that the element $a_{ij}$ of $\bm{A}$ indicates the existence of an edge going from node $j$ to node $i$.
In the directed case, the neighborhood of node $i$, i.e. the set of nodes that can be reached from $i$, is hence given by the non-zero entries on the $i$th column of $\bm{A}$, and the precursors of node $i$ are the set of nodes having a directed edge to node $i$.
The undirected case is obtained by placing a directed edge in both directions.

Following the prescription of Ref.~\cite{Gleeson2008}, all nodes are initially placed in the inactive state, except for a (potentially disconnected) set of seed nodes, initially active.
At each subsequent discrete time steps $t$, inactive nodes make a transition to the active state if their precursors satisfies an activation condition.
The process ends when no further transitions are possible (i.e., when the state of each node at time $t$ is identical to the state at time $t+1$).

The transition conditions are encoded in a set of $N$ so-called ``influence response functions'' $\{F_i(m)\}_{i=1,..,N}$, where $m$ is the number of active precursors of node $i$ \cite{Gleeson2008}.
We define these functions as the cumulative distribution function (CDF) of the probability $P_i(m)$ that node $i$ has a hidden activation threshold precisely equal to $m$ precursors:
\begin{equation}
    P_i(m)=F_i(m)-F_i(m-1)\;,
\end{equation}
or in other words, as the CDF of the probability that node $i$ will make a transition to the active state, as soon as $m$ of its precursors reach the active state.
For convenience, we also define the complementary probability 
\begin{equation}
    G_i(m) := 1-F_i(m)
\end{equation}
that node $i$ has a hidden threshold greater than $m$ active precursors---or equivalently, that it does not make a transition when it has $m$ active precursors.
Under this probabilistic description, the process ends when every inactive node at time $t$ fails to make a transition to the active state.

A complete specification of the dynamics thus consists of: the structure of the network, as specified by a $N\times N$ adjacency matrix $\bm{A}$; a set of seed nodes; and a response function $F_i(m)$ for each node.
Because our formalism allows it, we  will work with response functions specified on a node-to-node basis.
Note, however, that it is not uncommon to group nodes in coarser compartments.
In fact, response functions $F(m_i, \theta_i)$ that only depend on $m_i$ and some local parameter $\theta_i$---such as the degree $k_i$ of node $i$---are widely used \cite{Gleeson2013}: Many well-known dynamics can be recovered as a special case by choosing specific classes of response functions  \cite{Gleeson2008}.

\subsection{Special classes of response functions}
\label{subsec:special_cases}
Let us consider first bond percolation, where a node is part of an active component if at least one of its incoming edges percolates.
If one denotes by $m_i$ the number of in-edges, that can percolate, of node $i$, and by $p$ the percolation probability, then node $i$ remains inactive with probability $G(m_i,p) = (1-p)^{m_i}$.
The response function can therefore be written as the complementary probability
\begin{subequations}
\begin{equation}
  \label{EQ:ReponseFunctionBond}  
  F(m_i) = 1-(1-p)^{m_i}\;.
\end{equation}
For site percolation, where a node percolates with probability $p$ if it has at least a single active precursor, the response function is simply
\begin{equation}
  \label{EQ:ReponseFunctionSite}  
  F(m_i)= 
  \begin{cases}
      0,&  m_i=0\\
      p,  & m_i>0.
  \end{cases}
\end{equation}
In the Watts cascade model \cite{Watts2002}, a node of degree $k_i$ makes a transition to the active state when a critical fraction $r$ of its $k_i$ precursors reaches the active state, and the thresholds $r$ are drawn from a distribution, whose cumulative distribution is given by $C(r)$.
The response function for the resulting dynamics is therefore
\begin{equation}
  \label{EQ:ReponseFunctionWatts} 
  F(m_i, k_i)= C\left(m_i/k_i\right).
\end{equation}
\end{subequations}
These are but a few examples, derived in Ref.~\cite{Gleeson2008}, and reproduced here to showcase the generality of the cascade formalism.
Furthermore, there exists many more equivalencies, inherited from its reduction to the Watts threshold model [see Eq.~\eqref{EQ:ReponseFunctionWatts}]; the latter can be mapped, with the appropriate choice of thresholds, to site percolation, $k$-core percolation, \redcopy{diffusion percolation}~\cite{adler1988diffusion}, and the generalized epidemic process \cite{miller2016equivalence}.
Our method therefore provides systematic and exact solutions to a large class of dynamics.

\subsection{Exact recursive solution}
\label{subsec:recursive}
The outcome of a cascade dynamics on a network is a configuration of active and inactive nodes. 
We encode these configurations in binary vectors $\bm{l} = [l_1,\hdots,l_N]^\intercal$ of length $N$, where $l_i = 1$ if node $i$ is active in the configuration, and $0$ otherwise.
Thus, the vector $\bm{n} := [1,\hdots,1]^\intercal$ refers to the completely active configuration, and the number of active nodes in a configuration $\bm{l}$ is given by the square of its Euclidean norm, e.g., $|\bm{n}|^2=N$.

Cascade dynamics are, by definition, probabilistic processes. 
So fully solving the cascade amounts to computing the probability of every outcome.
Given a fixed initial configuration $\bm{l}_0$ and a fixed structure $\bm{A}$, we define  $Q(\bm{l};\bm{A},\bm{l}_0)$ as the probability of observing configuration $\bm{l}$ when the cascade stops.
A solution is therefore a distribution  $\mathcal{Q}=\{ Q(\bm{l};\bm{A},\bm{l}_0) \}_{\bm{l}}$ over all $\bm{l}\in\{0,1\}^N$, for a choice of response functions (not explicitly denoted, for the sake of clarity).
Even though there are exponentially many possible outcomes  ($|\mathcal{Q}|=2^N$), the calculation of $\mathcal{Q}$ is greatly simplified by the structure underlying the distribution.

Observe that when a cascade process ends, the nodes can be partitioned in two subsets defined by their activity status.
In particular, nodes in the inactive subset must have, by definition, a hidden threshold superior to the number of their active precursors.
The probability of observing any subset of inactive nodes can thus be written in terms of the complementary probabilities $\{G_i(m)\}$.
Specifically, since each activation event is independent, we can write the stopping probability of the cascade in configuration $\bm{l}$ as
\begin{subequations}
\label{EQ:iterative_equations}
\begin{equation}
  \label{EQ:iterative_equation_A}
  Q(\bm{l};\bm{A}_{\bm{u}}) = Q(\bm{l};\bm{A}_{\bm{l}}) \prod_{i=1}^N \bigl[G_i(m_i(\bm{l}))\bigr]^{u_i(1 - l_i)},\quad \bm{u}\geq \bm{l}\;,
\end{equation}
where $m_i(\bm{l})=\sum_{j} a_{ij}l_j$ is the number of active precursors of node $i$,  and where $\bm{u}\geq \bm{l}$ is an elementwise inequality $u_i\geq l_i\forall i$.
The matrix $\bm{A}_{\bm{l}}$ denotes the reduced adjacency matrix
\begin{equation*}
  \bm{A}_{\bm{l}} = \bm{L} \bm{A}\ \bm{L}\;,
\end{equation*}
where $\bm{L}=\textrm{diag}(\bm{l})$ is a $N\times N$ diagonal matrix, whose entries are given by the vector $\bm{l}$.
The reduced adjacency matrix $\bm{A}_{\bm{l}}$ is almost identical to $\bm{A}$, except that the values on the $j$\textsuperscript{th} row and column are set to zero if $l_j=0$. 
It encodes the structure of the subgraph induced by the set of active nodes in configuration $\bm{l}$. Thus, if $\bm{l}=\bm{n}$, then $\bm{A_n}=\bm{A}$.

With this in mind, Eq.~\eqref{EQ:iterative_equation_A} is interpreted as follows: It states that the probability that a cascade taking place on $\bm{A}_{\bm{u}}$ will stop at configuration $\bm{l}$ is equal to the probability $Q(\bm{l};\bm{A}_{\bm{l}})$ of reaching a completely active subgraph induced by $\bm{l}$, times the probability $\prod_{i=1}^N[G_i(m_i)]^{u_i(1-l_i)}$ that the cascade does not spread to the remaining $|\bm{u}-\bm{l}|^2$ inactive nodes.

This essentially solves the problem.
The distribution of outcomes $\mathcal{Q}$ can---in principle---be computed by solving the system of $2^{(N-|\bm{l}_0|^2)}$ equations defined in \eqref{EQ:iterative_equation_A}, recursively, for configurations with increasing numbers of active nodes.
However, it turns out that this system is not complete: Every completely active configuration in a reduced subgraph is associated to a non-informative  Eq.~\eqref{EQ:iterative_equation_A}, because $u_i=l_i=1$ $\forall i$ yields the tautology
\begin{equation*}
  Q(\bm{l};\bm{A}_{\bm{l}})=Q(\bm{l};\bm{A}_{\bm{l}})\;.
\end{equation*}
Hence, we are dealing with an underdetermined system of equations.
Fortunately, the system can be completed via the normalization condition
\begin{equation}
    \label{EQ:iterative_equation_B}
   Q(\bm{l};\bm{A}_{\bm{l}}) = 1 - \sum_{\redcopy{\bm{u}:\bm{u}<\bm{l}}}  Q(\bm{u};\bm{A}_{\bm{l}})\;,
\end{equation}
\redcopy{where $\bm{u}:\bm{u}<\bm{l}$ means that the sum is taken over all $\bm{u}$ respecting the elementwise inequality $u_i<l_i\forall i$.}~Eq.\eqref{EQ:iterative_equation_B} states that the probability of reaching the fully active configuration $\bm{l}$ on the reduced adjacency matrix $\bm{A}_{\bm{l}}$ is given by the complementary probability of reaching \emph{any other state} with fewer active nodes (on $\bm{A}_{\bm{l}}$). 
Of course, one also needs to give an initial condition of the form
\begin{equation}
    \label{EQ:iterative_equation_C}
    Q(\bm{l}_0;\bm{A}_{\bm{l}_0})=1, \quad Q(\bm{l};\bm{A}_{\bm{l}})=0\quad \forall\ \redcopy{\bm{l}\leq \bm{l}_0}\;,
\end{equation}
\end{subequations}
to complete the system.

Before considering the practical aspect of this formalism, we note that an analogous derivation leads to almost identical equations in the case of bond percolation \cite{Allard2012a}.
In fact, we show in Appendix~\ref{Sec:AppendixBondPercolation}, that upon substitution of the response function \eqref{EQ:ReponseFunctionBond} in the equations, one recovers the formalism derived in Ref.~ \cite{Allard2012a} for bond percolation.

\section{Enumeration algorithms}
\label{sec:numerical_methods}
\subsection{General recursive solution method}
\label{subsec:general_recursive_method}
In practice, and as stated above, Eqs.~\eqref{EQ:iterative_equations} must be solved recursively.
One must feed the results of Eq.~\eqref{EQ:iterative_equation_A} into  Eq.~\eqref{EQ:iterative_equation_B}, and back into  Eq.~\eqref{EQ:iterative_equation_A}, using the initial condition Eq.~\eqref{EQ:iterative_equation_C} as a starting point.
To make things more concrete, let us follow through with the first few steps of the computation. 

First, we compute the probabilities $Q(\bm{l}_0;\bm{A})$ using Eq.~\eqref{EQ:iterative_equation_A} and the initial condition Eq.~\eqref{EQ:iterative_equation_C}.
Then, we proceed to configurations with one more active node, i.e., configurations $\bm{l_i}$ such that $|\bm{l}_i|^2=|\bm{l}_0|^2+1$.
These probabilities, of the form $Q(\bm{l}_i;\bm{A})$, can be computed from Eq.~\eqref{EQ:iterative_equation_A}.
But in so doing, we will need to resort to the normalization condition [Eq.~\eqref{EQ:iterative_equation_B}]. 
Once the stopping probabilities of  all configurations with one extra active node are computed, it is only a matter of repeating the process for all configurations with one more active node (i.e., $\bm{l}_i$ such that $|\bm{l}_i|^2=|\bm{l}_0|^2+2$), progressing toward larger and larger configurations, all the way up to the complete configuration of $|\bm{n}|^2=N$ nodes.
This scheme obviously constructs the complete distribution $\mathcal{Q}$ (as well as a complete set of distributions $\mathcal{Q}'$ on every possible subgraph of $\bm{A}$).
We work out an explicit example for a small tadpole graph in  Appendix~\ref{Sec:AppendixExample}.

\subsection{Saving time: culling impossible configurations}
\label{sub:BFSRecursion}

\begin{figure}
    \centering
    \includegraphics[scale=0.25]{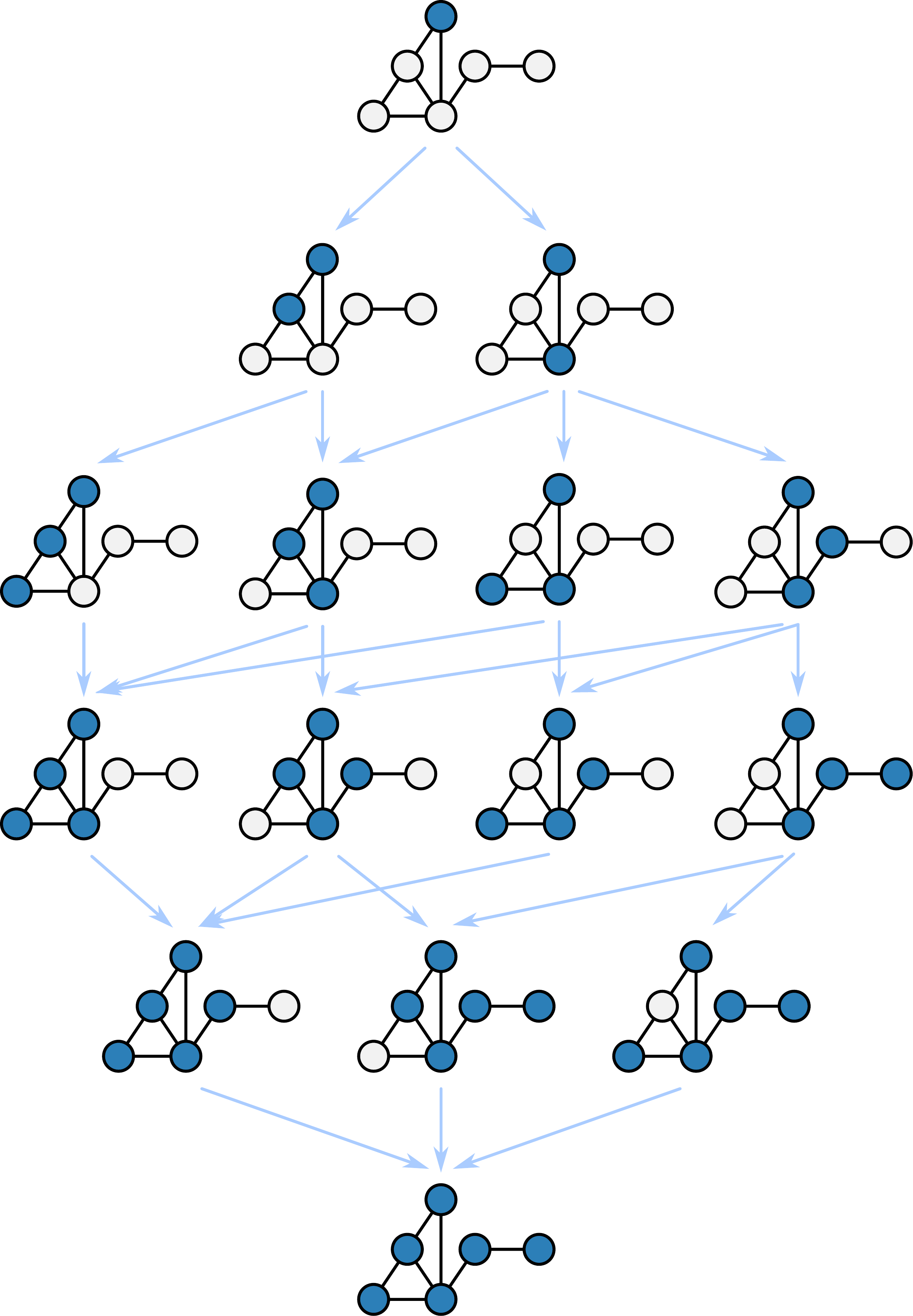}
    \caption{(Color online) Example of a breath-first exploration of all configurations. The configurations are explored by rows, i.e., by the number of discovered (active) nodes.
    Nodes are labeled as discovered (blue) or undiscovered (white). Blue arrows indicate possible transitions.} 
    \label{Fig:BreathFirst}
\end{figure} 

The method just discussed is less than optimal, precisely because it goes through every single configuration of active nodes and every single induced subgraphs.
This is due to the fact that many response functions and many graphs are associated with large sets of \emph{impossible} configurations, i.e., sets of configurations $\bm{l}$ with null probabilities $Q(\bm{l};\bm{A}_{\bm{u}})=0$.
It is easy to see that a configuration $\bm{l}$ is \emph{necessarily} impossible if it contains at least one node $i$ not connected to a seed (or a spontaneously activated node) via a path in $\bm{A}_{\bm{l}}$: There is then simply no possible path for the cascade to have propagated to node $i$.
Such configurations abound.

If we can find and avoid these impossible configurations more efficiently than through brute-force  enumeration of $\bm{l}\in\{0,1\}^N$, then we will have accelerated the calculation of $\mathcal{Q}$ by completely ignoring large portions of its support.
This can be done in most practical cases by using a breath-first search (BFS) \emph{over the configurations}, that accounts for the presence of spontaneously activated nodes [i.e. $F_i(0)>0$] and disconnected seeds, see Fig.~\ref{Fig:BreathFirst}.
In this modified BFS, nodes can be in one of two states: either discovered or undiscovered.
We seed the search with the initial condition $\bm{l}_0$, i.e., by labeling the  $|\bm{l}_0|^2$ seed nodes as discovered.
We then enumerate all \emph{neighboring} configurations containing $|\bm{l}_0|^2+1$ discovered nodes, i.e., all configurations  with $|\bm{l}_0|^2+1$ discovered nodes that are either neighbor of an initial seed node or a spontaneous active node .
We then repeat the process, expanding outwards.
Importantly, every new configuration is constructed by taking one of the previous configurations and converting one undiscovered node, adjacent to either an already discovered or a spontaneously activating nodes.

By carrying this process recursively, we mimic the cascade and only generate possible configurations.
When there are no undiscovered nodes adjacent to a discovered node and no undiscovered spontaneous nodes, we therefore have enumerated all possible final configurations, and the algorithm terminates. 

The advantage of using this search is that we can intertwine the enumeration procedure and the recursion of Eqs.~\eqref{EQ:iterative_equations}.
As soon as the number of active nodes increases by one in the BFS, say from $m$ to $m+1$, we know that we have encountered all possible active configurations with $m'\leq m$ nodes.
We can thus set the probability of all the missing configurations of less than $m+1$ nodes to 0.
At best, we will have saved vast amount of unnecessary calculation.
At worst, the complexity of the whole calculation will remain more or less the same---the BFS will only have ordered the enumeration of configurations.

\subsection{Saving more time: dropping intermediary distributions} 
\label{CompleteComputation}
%
\begin{figure*}[!t]
\centering
\includegraphics[trim = 0mm 2mm 0mm 0mm, clip=true, height=13\baselineskip]{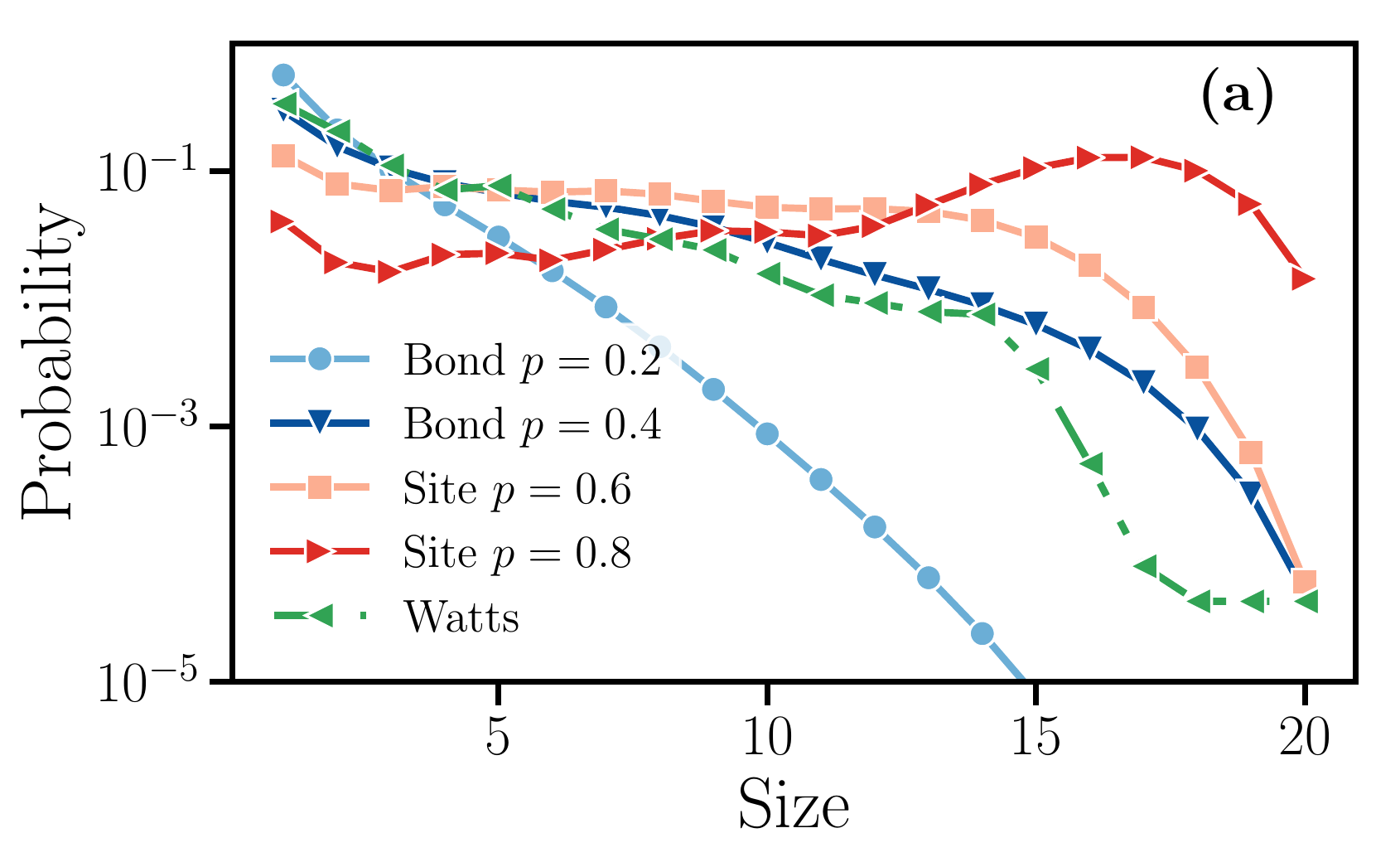}\qquad
\includegraphics[trim = 0mm 2mm 0mm 0mm, clip=true, height=13\baselineskip]{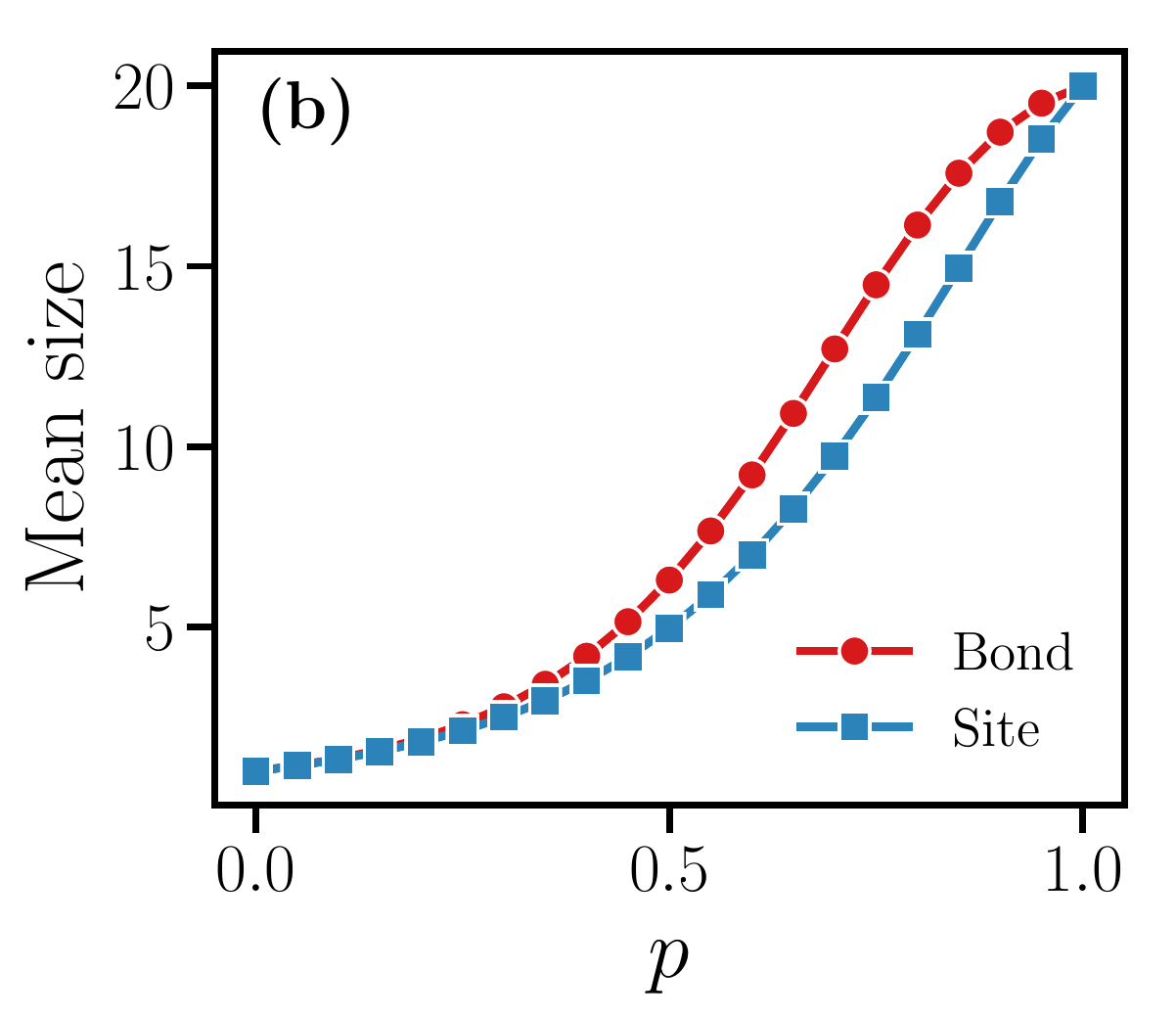}
\caption{(Color online) (a) Size distribution of the active components under different dynamic processes: bond percolation, site percolation and the Watts threshold model. Symbols are the results of $10^{7}$ Monte-Carlo simulations, and curves show the exact probability given by our approach~\redcopy{ Eq.\,\eqref{Eq:sizedistribution}}. (b) Exact mean size, \redcopy{Eq.\,\eqref{Eq:averageSize}}, of the final active component for bond and site percolations as a function of the occupation probability $p$. The graph has $N=20$ and 54 directed edges.}
\label{Fig:SizeDistriMeanGiant}
\end{figure*}
%
%
For most graphs, carrying out the full recursive solution of Eqs.~\eqref{EQ:iterative_equations} is impossible, because too much information must be  held either in memory, or because the calculation is simply too long---even with the help of the BFS strategy.

Besides the obvious exponential size of the solution, this slowdown can be traced to another important bottleneck: The evaluation of Eq. \eqref{EQ:iterative_equation_B} requires a complete knowledge of the distribution for all $\bm{u}$ with \redcopy{$\bm{u}<\bm{l}$}, i.e., all $\{\mathcal{Q}'\}$ over induced subgraphs.
In fact, one can show that $|\{\mathcal{Q}'\}|\sim N!$ (see Appendix.~\ref{Sec:AppendixComplexity}).
If we could do without the information contained in $\{\mathcal{Q}'\}$, then the limiting factor would become the (exponential) size of the solution $\mathcal{Q}$ rather the (factorial) complexity of the algorithm---an appreciable speedup.
This would of course come at the price of dropping the distributions $\{\mathcal{Q}'\}$ on smaller subgraphs.
One should therefore stick to the BFS enhanced recursion method of Sec.~\ref{sub:BFSRecursion} if this information is needed.

However, provided that $\{\mathcal{Q}'\}$ can be dropped, it is possible to compute $\mathcal{Q}$ directly.
The strategy consists in combining the recursive equations, and using BFS to enumerate the elements of the distribution $\mathcal{Q}$ in an orderly manner.
We build upon the observation that two configurations that share a common core of active nodes have related probabilities.
Specifically: the probability $Q(\bm{u};\bm{A_l})$ can be computed from the probability $Q(\bm{u};\bm{A})$ by considering the contribution of the inactive nodes absent from the subgraph $\bm{A_l}$:
\begin{subequations}
    \label{Eq:rewriteOnGraph}
\begin{equation}
    \label{Eq:rewriteOnGraphA}
    Q(\bm{u};\bm{A_l })=Q(\bm{u};\bm{A})\left[\prod_{i=1}^NG_i(m_i(\bm{u}))^{(1-l_i)(1-u_i)}\right]^{-1} .
\end{equation} 
[we merely ``factor out'' complementary probabilities hidden in $Q(\bm{u};\bm{A})$.]
Inserting Eq.~\eqref{Eq:rewriteOnGraphA} into Eq.~\eqref{EQ:iterative_equations}, we can then write the elements of $\mathcal{Q}$ directly as
  \begin{equation}
    \label{EQ:rewiteOnGraphB}
    Q(\bm{l};\bm{A})=\bigl[1 - Z(\bm{l},\bm{A})\bigr] \prod_{i=1}^N[G_i(m_i(\bm{l}))]^{(1-l_i)},
\end{equation}
where
\begin{equation}
    \label{EQ:rewiteOnGraphC}
  Z(\bm{l},\bm{A}):= \sum_{\redcopy{\bm{u}:\bm{u}<\bm{l}}} \frac{Q(\bm{u};\bm{A})}{\prod_{i=1}^NG_i(m_i(\bm{u}))^{(1-l_i)(1-u_i)}} \;.
\end{equation}
\end{subequations}

This transformation is useful because Eqs.~\eqref{EQ:rewiteOnGraphB}--\eqref{EQ:rewiteOnGraphC} explicitly include the normalization condition and can be solved without recursion.
The procedure is simple. 
We start by listing and ordering all possible configurations of the complete graph using BFS.
It is important to keep the discovery order.
Then, we solve Eq.~\eqref{EQ:rewiteOnGraphB} from the smallest configuration to the largest configuration (i.e. the discovery order). At each evaluation, every $Q(\bm{u};\bm{A})$ appearing in  Eq.~\eqref{EQ:rewiteOnGraphC} is already calculated and memorized---as imposed by the discovery order.
Under this systematic method, the previously calculated information is reinjected in the set of equations.
All intermediate distributions are thus simply dropped, which considerably reduces the computational complexity of the algorithm (see Appendix~\ref{Sec:AppendixComplexity}).

\section{Results and applications}
\label{sec:results}

Our main motivation in deriving recursive equations is to elaborate methods that combines exactly solved motifs on a tree-like backbone, in the spirit of Refs.~\cite{Allard2015,Karrer2010}.
However, for small graphs, the method can also be useful in itself.
For instance, one can marginalize $\mathcal{Q}$ to obtain the individual activation probabilities of each node, and these probabilities can then be used for diagnosis purposes, e.g., the importance of a node with regard to spreading \cite{holme2017three}.
With this in mind, we give two practical examples in the next section, to illustrate the power of exact solutions on small graphs.

\subsection{Calibration: special cases on a directed graph}
%
%
%
To calibrate the method and verify its validity, we first use the algorithm of Sec.~\ref{CompleteComputation} to obtain cascade results on a known network: The directed network of 20 nodes and 54 directed edges appearing in Ref.~\cite{Allard2012a}. We compute the solutions for three special cases of cascade dynamics, previously introduced in Sec.~\ref{subsec:special_cases}: bond percolation [see Eq.~\eqref{EQ:ReponseFunctionBond}], site percolation [see Eq.~\eqref{EQ:ReponseFunctionSite}], and the Watts cascade model [see Eq.~\eqref{EQ:ReponseFunctionWatts}].
Furthermore, we compare the predictions of the formalism with Monte-Carlo simulations. 

For site (bond) percolation, the Monte-Carlo simulations are done by randomly selecting a seed node and then occupying the neighbors (adjacent edges) with a probability $p$.
This step is then repeated for the new neighborhood (adjacent edges), until the process stops or until the network is exhausted.
For the Watts threshold model, the simulation follows the standard description of a cascade process \cite{Watts2002}.
We first assign a threshold $C_i$ between 0 and 1 to each node $i$,  drawn from a uniform distribution.
A randomly chosen node---the seed---is then activated.
We then enter the propagation phase: If the threshold $C_i$ of node $i$ is lower or equal to the fraction of its precursors that are active $m_i/k_i$, then the node makes a transition to the active state.
This process is repeated until no transitions are possible.
For all these Monte-Carlo simulations, the estimator $\hat{Q}(\bm{l};\bm{A},\bm{l}_0)$ of the outcome probability is computed as the frequency of the final configuration $\bm{l}$; standard results tell us that the variance of $\hat{Q}$ will decrease as the square root of the number of trials.

Note that because we seed Monte-Carlo simulations at random,  we compare against results computed using exact probabilities $\mathcal{Q}$ that are marginalized over every initial configuration $|\bm{l}_0|^2=1$, i.e.,
\begin{equation*}
  Q(\bm{l};\bm{A}) =  \sum_{\bm{l}_0 : |\bm{l}_0|^2=1} Q(\bm{l};\bm{A},\bm{l}_0)\;.
\end{equation*}
This is consistent with the typical way in which Monte-Carlo simulations are carried out.

The graph contains $20$ nodes, meaning that there are roughly $10^6$ different configurations---we cannot possibly visualize the complete distribution $\mathcal{Q}$.
Thus, we will focus on summary statistics, the size distribution [Fig.~\ref{Fig:SizeDistriMeanGiant}~(a)] and the mean size [Fig.~\ref{Fig:SizeDistriMeanGiant}~(b)].
Using the distribution $\mathcal{Q}$, we calculate the probability $p_s$ that a cascade will reach $s$ nodes as 
\begin{equation}
  p_s(\bm{A}) = \sum_{\bm{l}\in\{0,1\}^N} Q(\bm{l};\bm{A}) \mathbb{I}(|\bm{l}|^2=s)\;,\label{Eq:sizedistribution}
\end{equation}
where $\mathbb{I}(\cdot)$ is an indicator function, equal to 1 when its argument is true, and 0 otherwise.
We also compute the mean size of the active component 
\begin{equation}
  \redcopy{S(\bm{A}) = \sum_{s=1}^N  sp_s(\bm{A}) = \sum_{\{\bm{l}\}} Q(\bm{l};\bm{A})|\bm{l}|^2\;.}\label{Eq:averageSize}
\end{equation}

Figure~\ref{Fig:SizeDistriMeanGiant}~(a) shows the size distribution for different dynamics and occupation probabilities.
The perfect fit of the exact results and the Monte-Carlo simulations confirms the validity the equations and the algorithm.
In the case of the selected graph (see Ref.~\cite{Allard2012a}), roughly half of the $2^{20}=1\,048\,576$ possible configurations go into the calculations of the size distributions.
Notably, computing these distributions is trivial once the recursive procedure of Sec.~\ref{sec:numerical_methods} has been carried out: The equations yield a large polynomial in $\{G_i(x)\}$, which can be evaluated easily and many times, by specifying the response functions and parameters.
Thus, generating a family of distributions is not significantly more costly than generating a single one.
Another important aspect, also raised in Ref.\cite{Allard2012a} is the irregularities of the size distributions (in the case of bond percolation), which indicates that the solution is non-trivial and depends on the intricacies of the structure of the graph.
This tells us that any close form solution will necessarily be just as complex.
Furthermore, we see in Fig.~\ref{Fig:SizeDistriMeanGiant}~(a) that bond percolation behaves, in fact, quite simply in comparison with the other dynamics (site percolation, Watts model); exact solutions on small graphs therefore appear even more useful in the case of general cascade dynamics than in the special case of bond percolation.

Figure~\ref{Fig:SizeDistriMeanGiant}~(b) shows the theoretical mean size component for two dynamics.
Again, we benefit from the fact that the corresponding polynomial can be evaluated at will once the recursive procedure has been completed: Our approach allows us to zoom-in on any point of the curve, and to investigate specific regions of the parameter space at little extra computational costs.

\subsection{Mixed dynamics } 

\begin{figure}
\includegraphics[trim = 0mm 0mm 0mm 0mm, clip, width=0.7\linewidth]{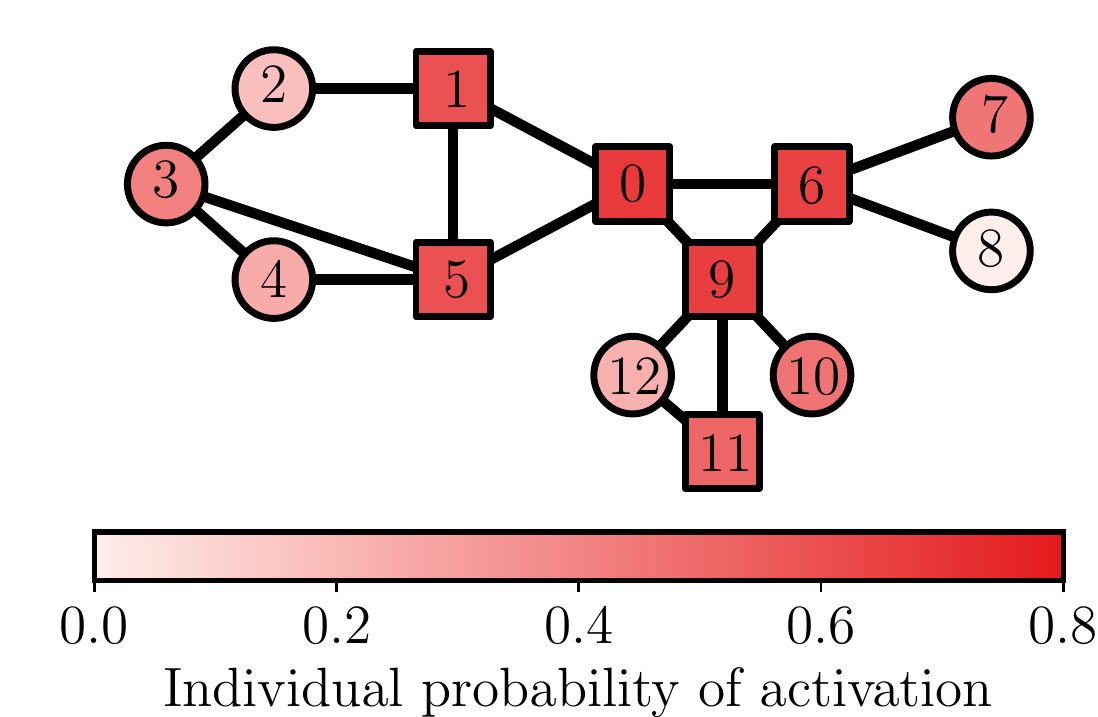}
\caption{(Color online) Individual response function and activation probabilities. The colors of the nodes are used to show their activation probability, summed over all possible outcomes. 
\redcopy{Square nodes can make a spontaneous transition [with response function $F_s(p,q)$, see Eq.~\eqref{eq:examplersponse_spont}], while round nodes need active precursors to make a transition [they have the response function $F_w(p,\tau)$, see Eq.~\eqref{eq:examplersponse_thresh}].}
The parameters of the spontaneously activating nodes are: $p=0.4$ and $q=0.3$ for nodes 1, 5 and 11; and $p=0.6$ and $q=0.1$ for nodes 0, 6 and 9.
The parameters of the threshold nodes are:  $p=0.6$ and $\tau=2$ for node 2, 4, 8 and 12; and $p=0.7$ and $\tau=1$, for nodes 3, 7 and 10.}
\label{Fig:Networks}
\end{figure}
\begin{figure}
\includegraphics[trim = 0mm 0mm 0mm 0mm, height=13\baselineskip]{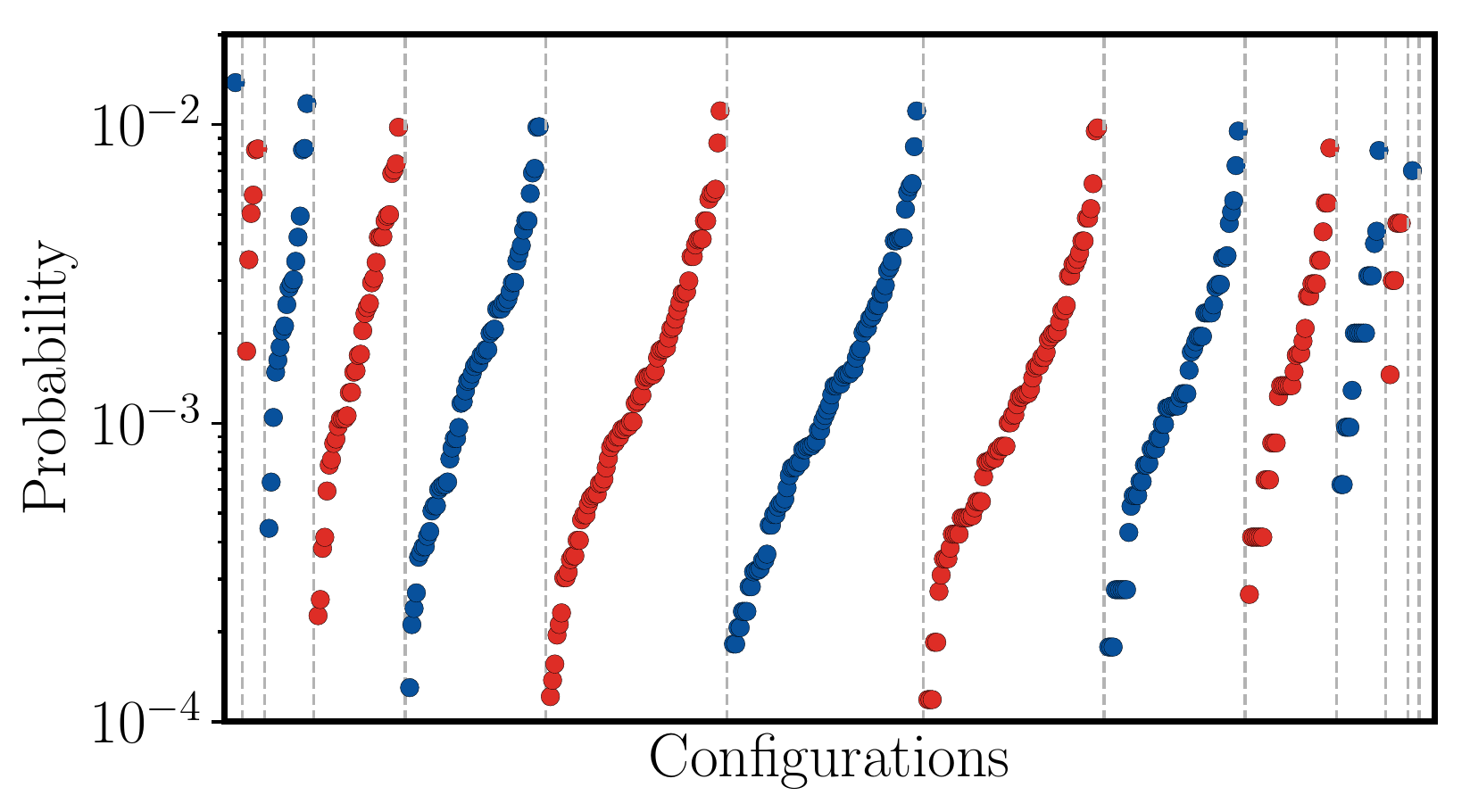}\qquad
\caption{(Color online) Complete distribution $\mathcal{Q}$ for the every possible configurations of the network, using the dynamics described in Fig.~\ref{Fig:Networks}.
Configurations are grouped together according to the number of active nodes (from 0 to 12), and then ordered with ascending probabilities. Since node $8$ is always inactive, the completely active configuration ($13$ active nodes) has a zero probability and, consequently, is not shown.
Gray lines are placed where the number of active nodes increases.}
\label{Fig:Individual}
\end{figure}


In the previous case study, the network had a global response function, independent of the node identity.
In real situations, it is fair to assume that nodes of different types do not respond identically to stimuli.
Thus, it is natural and perfectly general to consider a mixed dynamics that stems from diverse response functions.
As highlighted in Sec.~\ref{sec:formalism}, our formalism can handle this generalization  without any additional complexity.

We consider a mixed dynamics where nodes are associated with one of two types of parameterizable response function, 
\begin{align}
 &F_{w}(p,\tau)= \left\{\begin{array}{ll}
  0 & m\leq \tau-1\;,\\
  p & m\geq \tau \;,  
 \end{array}\right.\label{eq:examplersponse_thresh}\\
 &F_{s}(p,q)= \left\{\begin{array}{ll}
  p & m = 0 \;,\\
  1 - (1 - p)q^m & \redcopy{m> 0}   .
 \end{array}\right.\label{eq:examplersponse_spont}
\end{align}
The function $F_w(p,\tau)$ describes a threshold dynamics, where $\tau$ is the activation threshold and $p$ is the activation probability once the threshold is exceeded.
The function $F_s(p,q)$ describes a node that can either activate spontaneously, or through contagion; $p$ is the probability that the node will spontaneously activate (at time $t=0$), whereas $q$ controls its sensitivity to active precursors.
We assign these response functions to the $N=13$ nodes of a handmade small network, shown in Fig.~\ref{Fig:Networks}. \redcopy{The network is constructed to showcase heterogeneous patterns of activation.}

Because we have introduced random seeds, we will no longer sum the probabilities over all seeds; instead, we always begin the process in the inactive configuration $\bm{l_0}=[0,0,\hdots,0]\intercal$ and let spontaneous activations determine the outcome.
In Fig.~\ref{Fig:Individual}, the probability $Q(\bm{l};\bm{A})$ is displayed for the complete set of possible outcomes $\bm{l}$.
Since the network is relatively sparse and small ($N=13$), these results can be computed extremely quickly (less than a minute on a modern personal computer).

In Fig.~\ref{Fig:Networks}, we have colored nodes with their probability of being active by summing over all possible outcomes.
Combining the information given by both of these graphs, we see that the fully active configuration is impossible, because nodes 8 has a degree smaller than its threshold. Similar and much more in depth analyses could obviously be carried out on real small networks, e.g., on a power grid.

\section{Discussion}
\label{sec:conclusion}
In this paper, we have derived a new set of recursive equations that solve cascade dynamics on arbitrary networks, and we have introduced two practical strategies to manage the complexity of solving such a large set of equations.
These developments are inspired by analogous methods, introduced within the framework of percolation theory \cite{Allard2012a}.

Due to the generality of the cascade dynamics formulation \cite{Gleeson2008,miller2016equivalence}, our method leads to exact solutions for a wide range of dynamics, including well-known examples such as site and bond percolation and the Watts threshold model. 
But its power goes beyond such simple cases; it also generates exact solution for many exotic dynamics and connectivity patterns, e.g., directed, self-referencing edges, weighted graphs, disconnected active components, spontaneous activity and disconnected initial seeds activation. \redcopy{Furthermore, the exact solution is valid in the much general context of individual activation functions (i.e., each node can have a different activation function). To the best of our knowledge, our framework is the only one able to do so.}

We cannot understate the fact that our formalism solves a problem whose solution grows exponentially with $N$. 
As such, the method is, by necessity, intractable\footnote{
In the worst case scenario of a complete graph, our implementation is able to handle roughly 20 nodes on a modern single-CPU computer.}. \redcopy{On the other hand, the usual methods for solving numerically cascade dynamics, the tree-based theory and the message-passing framework \cite{lokhov2015dynamic,gleeson2017message,Karrer2014}, are surprisingly accurate for sparse networks and are commonly used on real large networks \cite{Melnik2011}.} Our contribution is therefore not designed for computing ensemble statistics in the large $N$ limit, where Monte-Carlo simluations or approximations schemes would be more appropriate, if the specifics of the configurations do not matter.
Instead, our method is designed to yield exact solutions for small graphs, where the exponential dependency is still manageable.
This is useful in at least three scenarios: (i) on small graphs where accurate \emph{configuration} probabilities are needed (c.f. Ref.~\cite{holme2017three}), (ii) in formalisms where motif distributions are specified and traversal probabilities must preferably be computed in closed form (c.f. Refs.~\cite{Allard2015,Allard2012a,Allard2012b}), \redcopy{ and (iii) on graphs with mixed dynamics. } 
In all cases, our formalism involves calculations that are no more complex than Monte-Carlo simulations, with the added advantage of producing exact probabilities as well.

In summary, despite the unwieldiness of the calculations involved, our results open the way to new theoretical predictions, since it solves cascades on small motifs, both exactly and systematically.

\section*{Acknowledgments}
We are thankful to Antoine Allard and Patrick Desrosiers for useful comments and suggestions. We are grateful to Andrey Lokhov for pointing out Ref.~\cite{lokhov2015dynamic}.
This work was funded by the Fonds de recherche du Qu\'ebec-Nature et technologies (J.-G.Y and E.L.), the Conseil de recherches en sciences naturelles et en g\'enie du Canada (L.J.D.), the Irish Research Council (New Foundations grant to S.M.), Science Foundation Ireland (Grant No. 11/PI/1026, S.M.) and Sentinel North (E.L., J.-G.Y).\\
E.L. and J.-G.Y. contributed equally to this work.

\appendix
\section{Equivalence with bond percolation}
\label{Sec:AppendixBondPercolation}
An exact solution to bond percolation is given in Ref.~\cite{Allard2012a}, in the form of a set of recursive equations.
We show that our equations generalize these equations.
As stated in Sec.~\ref{subsec:special_cases}, bond percolation can be mapped to a cascade dynamic process by setting $F(m_i)=1-(1-p)^{m_i}$ and $G(m_i)=(1-p)^{m_i}$ for all nodes.
We insert these relations in Eqs.~\eqref{EQ:iterative_equation_A} and obtain
\begin{equation}
    \label{EQ:bondIterativeA}
    Q(\bm{l};\bm{A_u})=Q(\bm{l};\bm{A_l})\prod_{i=1}^N[(1-p)^{m_i}]^{u_i-l_i} \qquad \bm{u}\geq\bm{l}
\end{equation}
Next, writing out the number of active precursors of $i$ as $m_i(\bm{l})=\sum_k a_{ik}l_k$, we get 
\begin{align*}
  \prod_{i=1}^N (1-p)^{\sum_k a_{ik}l_k\bar{l}_i} & = 
   (1-p)^{\sum_{k,i} a_{ik}l_k\bar{l}_i}\;, \\
   &=(1-p)^{\bm{l}^T\bm{A}\bar{\bm{l}}}\;,
\end{align*}
where $\bar{l}_i:=u_i-l_i$.
Substituting back into Eq.~\eqref{EQ:bondIterativeA}, we find, for $\bm{u}=\bm{n}$,
\begin{equation}
    \label{EQ:bondIterativeB}
    Q(\bm{l};\bm{A})=Q(\bm{l};\bm{A_l})(1-p)^{\bm{l}^T\bm{A}\bar{\bm{l}}},
\end{equation}
i.e., Eq.~(3) of Ref.~\cite{Allard2012a}. The normalization condition is simply generalized since it does not depend on the specifics of the cascade dynamics.
This completes the proof of equivalence.

\section{Explicit example on a tadpole graph}
\label{Sec:AppendixExample}
In this Appendix, we work out several of the first steps of the full recursive procedure, on a small tadpole graph of 4 nodes, i.e., the graph whose adjacency matrix is given by
\begin{equation}\bm{A}=
  \begin{pmatrix}
   0 & 1 & 1 & 0 \\
   1 & 0 & 1 & 0 \\
   1 & 1 & 0 & 1 \\
   0 & 0 & 1 & 0 \\
\end{pmatrix}.
\end{equation}
We choose an initial configuration where every node is inactive, $\bm{l_0}=[0,0,0,0]^\intercal$.
To keep the response function as general as possible, we will not further specify the dynamics.
Therefore, spontaneous activation [$F_i(m)>0$] is assumed to occur, otherwise the only possible outcome is $\bm{l_0}$.
To simplify the demonstration, we suppose identical response functions for each node, i.e., $F_i(m)=F(m_i)~\forall~i$.

There exists $2^4$ different configurations of the cascade dynamics.
We start with the initial configuration and Eq.~\eqref{EQ:iterative_equation_A}, to yield
\begin{align*}
  Q(\bm{l_0};\bm{A}) &= Q(\bm{l_0};\bm{A_{l_0}})G(0)^4=G(0)^4,
\end{align*}
since $Q(\bm{l_0};\bm{A_{l_0}})=1$ by definition [see the initial condition in Eq.~\eqref{EQ:iterative_equation_C}].
Moving to configurations with one more active node, we begin with $\bm{l_1}=[1,0,0,0]^\intercal$ and follow the same procedure 
\begin{align}
   Q(\bm{l_1};\bm{A}) &= Q(\bm{l_1};\bm{A_{l_1}})G(1)^2G(0)\nonumber\\
    &=[1-Q(\bm{l_0};\bm{A_{l_1}})]G(1)^2G(0)\nonumber\\
    &=[1-Q(\bm{l_0};\bm{A_{l_0}})G(0)]G(1)^2G(0)\nonumber\\
    &=F(0)G(1)^2G(0), \label{eq:appendix_example_l1}
\end{align}
where we have used the definition $F(m)=1-G(m)$ at the last step. 
Notice how the normalization \eqref{EQ:iterative_equation_C} intervenes and how $Q(\bm{l_0};\bm{A_{l_1}})$ and $ Q(\bm{l_1};\bm{A_{l_1}})$ are computed as a by-product of this step, producing the complete distribution of outcomes $\mathcal{Q}'$ for a cascade taking place on $\bm{A_{l_1}}$.
Further steps will generate similar distributions.

Completing our calculations for the other 1 node configurations, we observe that the symmetry of the graph leads to $Q(\bm{l_2};\bm{A})=Q(\bm{l_1};\bm{A})=F(0)G(1)^2G(0)$ with $\bm{l_2}=[0,1,0,0]^\intercal$, while
\begin{align*}
  Q(\bm{l_3}=[0,0,1,0]^\intercal;\bm{A}) &= F(0)G(1)^3\;,\\
  Q(\bm{l_4}=[0,0,0,1]^\intercal;\bm{A}) &= F(0)G(1)G(0)^2\;,
\end{align*}
following the procedure of  Eq.~\eqref{eq:appendix_example_l1}. 

For  configurations with two active nodes  such as $\bm{l_5}=[1,1,0,0]^\intercal$, two steps of recursions are required.
First, we use Eq.~\eqref{EQ:iterative_equation_A}
 \begin{align}
   Q(\bm{l_5};\bm{A}) &= Q(\bm{l_5};\bm{A_{l_5}})G(2)G(0)\nonumber,
\end{align} 
and apply Eq.~\eqref{EQ:iterative_equation_B}
\begin{align*}
  Q(\bm{l_5};\bm{A_{l_5}}) = [1-Q(\bm{l_0};\bm{A_{l_5}})-Q(\bm{l_1};\bm{A_{l_5}})-Q(\bm{l_2};\bm{A_{l_5}})].
\end{align*}
None of these terms are a priori known. We must use Eq.~\eqref{EQ:iterative_equation_A} again for each of them 
\begin{align*}
  Q(\bm{l_0};\bm{A_{l_5}})&= Q(\bm{l_0};\bm{A_{l_0}})G(0)^2=G(0)^2,\\
  Q(\bm{l_1};\bm{A_{l_5}})&= Q(\bm{l_1};\bm{A_{l_1}})G(1)=F(1)G(1),\\
  Q(\bm{l_2};\bm{A_{l_5}})&= Q(\bm{l_2};\bm{A_{l_2}})G(1)=F(1)G(1).
\end{align*}
This leads to 
 \begin{align}
   Q(\bm{l_5};\bm{A}) &= [1-G(0)^2-2F(1)G(1)]G(2)G(0)\nonumber.
\end{align}
This process can obviously be carried out systematically for the 5 other configurations with 2 active nodes.
We then proceed to larger configurations until we reach $\bm{l}=\bm{n}$, which requires a special treatment.
Rather than using Eq.~\eqref{EQ:iterative_equation_A}, we directly use the normalization \eqref{EQ:iterative_equation_B} to find $Q(\bm{n};\bm{A})$.
This completes the calculation of $\mathcal{Q}$ on $\bm{A}$, and all other distributions $\mathcal{Q}'<\mathcal{Q}$ have been computed in the process.

\begin{figure}
    \centering 
    \includegraphics[width=0.65\linewidth]{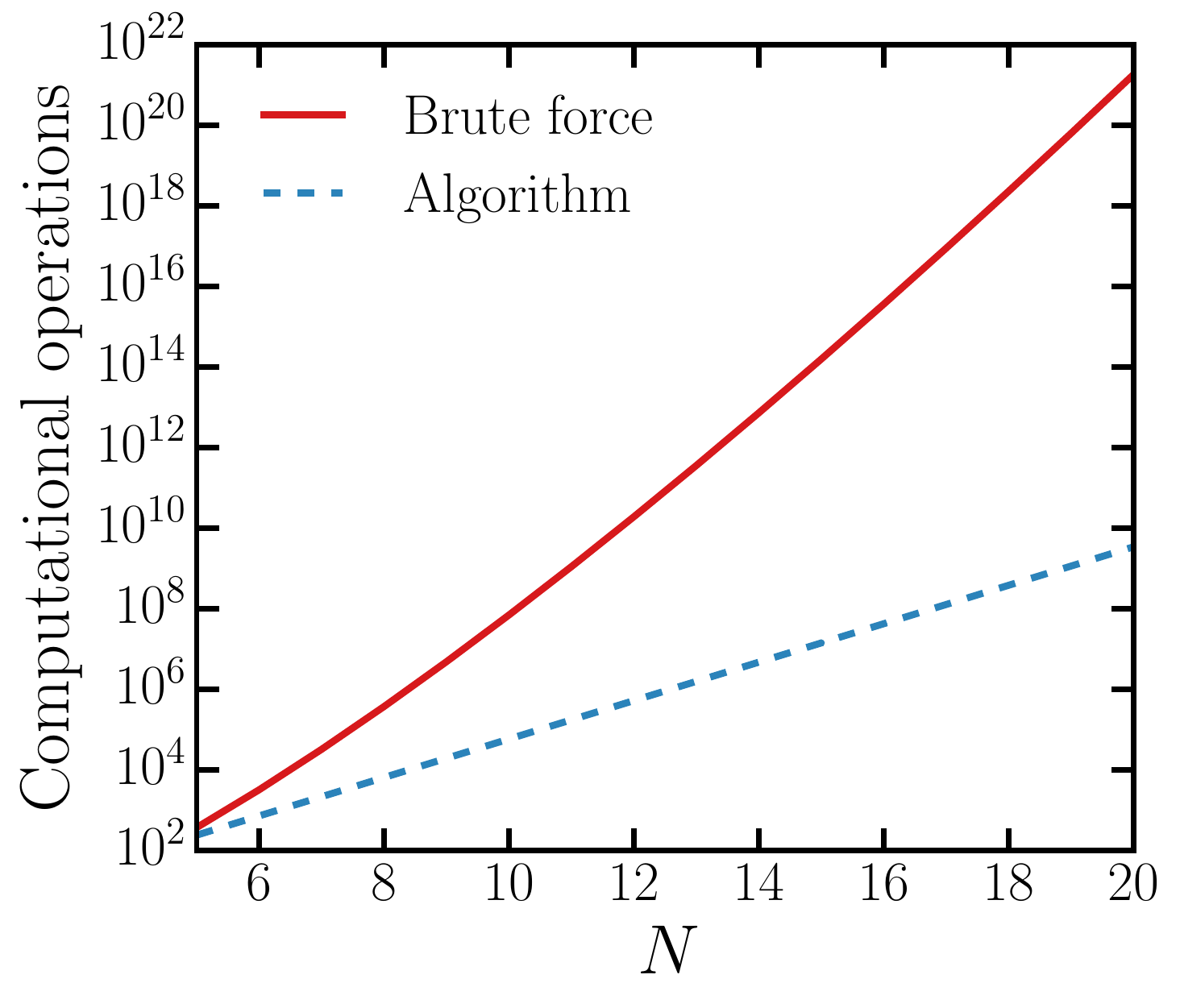}
    \caption{Worst-case complexity of a brute-force solution of the Eqs.~\eqref{EQ:iterative_equations}, and of the accelerated solution Eq.~\eqref{EQ:rewiteOnGraphB}.
    The complexity is assumed to be proportional to the number of $Q(\bm{l};\bm{A_u})$ required to obtain a solution. The worst-case corresponds to an arbitrary dynamics with spontaneous activation on  a complete graph of $N$ nodes.
    }
      \label{Fig:Complexity}
\end{figure}  

\section{Complexity calculation}
\label{Sec:AppendixComplexity}
In the main text, we mention that the exact recursive solution of Eqs.~\eqref{EQ:iterative_equations} is much more computationally complex than the accelerated solution, found using Eqs.~\eqref{EQ:rewiteOnGraphB}, since the latter skips the computation of $\mathcal{Q}'$ on the induced subgraphs.
This Appendix clarifies and quantifies this statement.
We consider the worst case: An undirected complete graph of $N$ nodes, with an undefined dynamics and no initial activation, i.e., $|\bm{l_0}|^2=0$.
In this setup, the number of possible configurations is $2^N$ and the BFS is unnecessary.
We assume that the complexity of the process is well represented by the number of $Q(\bm{l};\bm{A_u})$ required to obtain a solution.  

We begin with the analysis of the complexity of the algorithm of Sec.~\ref{CompleteComputation} (the accelerated solution).
For a configuration $\bm{l}$, we count the number of terms $\mathcal{N}_{|\bm{l}|^2}$ involved in the evaluation of $Q(\bm{l};\bm{A})$ from Eq.~\eqref{EQ:rewiteOnGraphB}.
\begin{equation}
  \mathcal{N}_{|\bm{l}|^2}=O\left(\sum_{|\bm{u}|<|\bm{l}|} Q(\bm{u};\bm{A_n })\right).
\end{equation}
For each $|\bm{u}|^2$ of the sum, there exists $\binom{|\bm{l}|^2}{|\bm{u}|^2}$ different configurations, meaning that
\begin{equation}
  \mathcal{N}_{|\bm{l}|^2}=\sum_{|\bm{u}|^2=0}^{|\bm{l}|^2-1}\binom{|\bm{l}|^2}{|\bm{u}|^2}\sim 2^{|\bm{l}|^2}
\end{equation}
For a configuration $\bm{l}$, roughly $2^{|\bm{l}|^2}$ terms are needed to obtain a solution to Eq.~\eqref{EQ:rewiteOnGraphB}.
To obtain the complexity of the complete distribution $\lbrace Q(\bm{l};\bm{A})\rbrace_{\bm{l}}$, we sum up the complexity of all $2^{N}$ equations,
\begin{align}
  \mathcal{N}_{\text{total}} &=\sum_{\bm{l}}\mathcal{N}_{|\bm{l}|^2} \\
  &=\sum_{|\bm{l}|^2=0}^{N}\binom{N}{|\bm{l}|^2}2^{|\bm{l}|^2} = 3^N\label{Eq:ComplexityAlgorithm}.
\end{align}
We conclude that the total number of operations scales approximately as $3^N$ for a complete set of configurations for an arbitrary dynamics on a complete graph, using the algorithm of Sec.~\ref{CompleteComputation}.

Next, we analyze the complexity of the naive recursion algorithm (see Sec.~\ref{subsec:general_recursive_method}) in the same manner.
We call this method the brute-force method, since it does not use any shortcuts.
Again, the complexity is represented by the number of $Q(\bm{l};\bm{A_u})$ required to obtain a solution.
The dominant contribution to the complexity comes from Eq.~\eqref{EQ:iterative_equation_B}, i.e., the normalization which needs to be evaluated for every $\mathcal{Q}'$ on all the induced subgraphs.
We denote by $\mathcal{M}_{|\bm{l}|^2}$ the complexity of the brute force calculation of $Q(\bm{l};\bm{A_u})$.
The complexity can be written as a recursive equation:
\begin{equation}
  \mathcal{M}_{|\bm{l}|^2} = O\left( \sum_{|\bm{u}|<|\bm{l}|} Q(\bm{u};\bm{A_l}) \right),
  \label{Eq:recursivecomplexity}
\end{equation}
\begin{align}
  \mathcal{M}_{|\bm{l}|^2} = \sum_{|\bm{u}|^2=0}^{|\bm{l}|^2-1}\mathcal{M}_{|\bm{u}|^2}\binom{|\bm{l}|^2}{|\bm{u}|^2},
\end{align}
since the normalization must be invoked all the way down to the initial condition.
Finally, summing $\mathcal{M}_{|\bm{l}|^2}$ over each configuration to obtain the total complexity for a complete set of outcomes, we find
\begin{align}
  \mathcal{M}_{\text{total}} =\sum_{|\bm{l}|^2=0}^{N}\binom{N}{|\bm{l}|^2}\mathcal{M}_{|\bm{l}|^2} \label{Eq:ComplexityBrute}.
\end{align}
For large $|\bm{l}|^2$, the last element of the series, i.e., $|\bm{u}|^2=|\bm{l}|^2-1$, is dominant and $\mathcal{M}_{|\bm{l}|^2}\sim |\bm{l}|^2!$ (see Eq.~\eqref{Eq:recursivecomplexity}).
Thus, $\mathcal{M}_{\text{total}}\sim N!$ for large $N$.

Figure \ref{Fig:Complexity} compares Eqs.~\eqref{Eq:ComplexityAlgorithm} and \eqref{Eq:ComplexityBrute} for different graph sizes.
It is clear that the brute-force method becomes impractically \emph{much faster} than the accelerated method.

%

\end{document}